\documentclass[electronic]{vgtc}             

\graphicspath{{figures/}{pictures/}{images/}{./}} 

\usepackage{times}                     

\usepackage{xspace}
\newcommand{\sm}{\emph{Simulacra Naturae}\xspace}

\newcommand{\ch}[1]{#1}

\usepackage{mathptmx}                  

\onlineid{1116}

\vgtccategory{Research}

\vgtcinsertpkg

\title{\sm: Generative Ecosystem driven by Agent-Based Simulations and Brain Organoid Collective Intelligence}

\author{
    Nefeli Manoudaki\thanks{nefeli@ucsb.edu; First three authors contributed equally.} \\ %
     \parbox{1.4in} {\scriptsize \centering Media Arts \& Technology \\ UC Santa Barbara}
\and Mert Toka\thanks{merttoka@ucsb.edu}\\ %
     \parbox{1.4in} {\scriptsize \centering Media Arts \& Technology \\ UC Santa Barbara}
\and Iason Paterakis\thanks{iason@ucsb.edu}\\ %
     \parbox{1.4in} {\scriptsize \centering Media Arts \& Technology \\ UC Santa Barbara}
\and Diarmid Flatley\thanks{diarmid@ucsb.edu}\\ %
     \parbox{1.4in} {\scriptsize \centering Media Arts \& Technology \\ UC Santa Barbara}
}

\teaser{
  \centering
  \includegraphics[width=\linewidth]{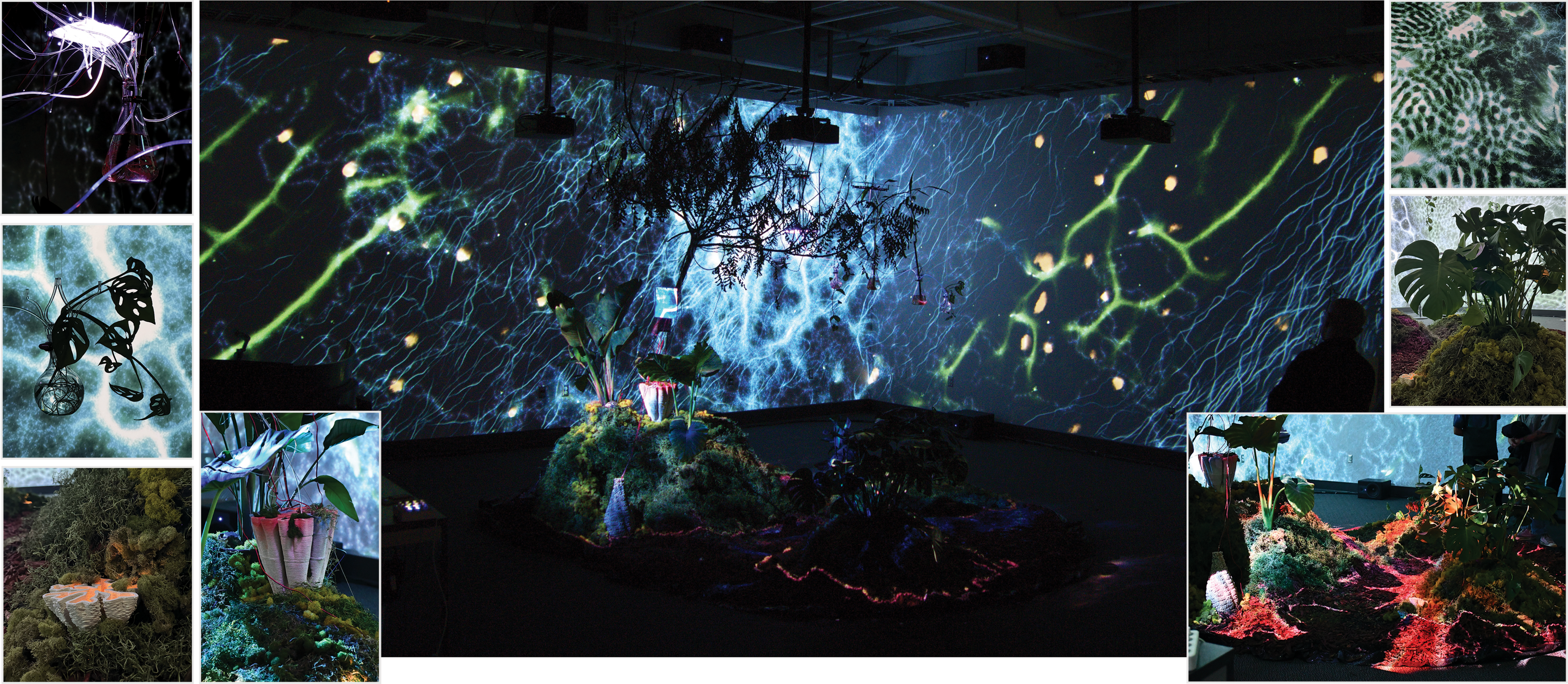}
  \caption{The installation of \sm. (Left) Close-up images of (top-to-bottom) the LED matrix; hydroponic plant in glassware; two morphogenic ceramic vessels. (Right) Detailed photos of (top-to-bottom) a variation of the projected visuals; one of the islands in the forest-like environment; floor projection on the forest-like environment.}
  \label{fig:teaser}
}

\abstract{
\sm is a data-driven media installation that explores collective care through the entanglement of biological computation, material ecologies, and generative systems. The work translates pre-recorded neural activity from brain organoids, lab-grown three-dimensional clusters of neurons, into a multi-sensory environment composed of generative visuals, spatial audio, living plants, and fabricated clay artifacts. 
These biosignals, streamed through a real-time system, modulate emergent agent behaviors inspired by natural systems such as termite colonies and slime molds. Rather than using biosignals as direct control inputs, \sm treats organoid activity as a co-creative force, allowing neural rhythms to guide the growth, form, and atmosphere of a generative ecosystem. The installation features computationally fabricated clay prints embedded with solenoids, \ch{adding physical sound resonances to the generative surround composition.} The spatial environment, filled with live tropical plants \ch{and a floor-level projection layer featuring real-time generative AI visuals}, invites participants into a sensory field shaped by nonhuman cognition. 
By grounding abstract data in living materials and embodied experience, \sm reimagines visualization as a practice of care, one that decentralizes human agency and opens new spaces for ethics, empathy, and ecological attunement within hybrid computational systems.

} 


\begin{document}


\firstsection{Introduction}
\maketitle

Recent advances in bio-art and data-driven installation practice have expanded the scope of visualization research beyond screen-based media toward multisensory, spatial experiences.  Projects such as Kac's \textit{Genesis} and Penny's responsive environments foreground living matter and distributed computation as aesthetic media~\cite{kac2007SignsLifeBio,penny2015EmergenceAgencyInteraction}.  Yet few works leverage \textit{brain-organoid} data (three-dimensional neural cultures that exhibit spontaneous, self-organizing activity) as a generative driver for audiovisual ecologies. This absence is surprising, because organoid electrophysiology offers rich, high-dimensional time series that align conceptually with the IEEE VIS community's interest in complex, dynamic data.

\sm addresses this gap by translating 131-channel organoid spike recordings into a multi-species agent simulation, a 16.2-channel spatial soundscape, a forest-like arrangement of live tropical plants, and a set of solenoid-actuated ceramic vessels.  The resulting 9m x 6m environment renders neural rhythms as evolving visual, acoustic, olfactory, and tactile phenomena, inviting audiences to experience data not as an abstract plot but as an inhabitable ecosystem.

The title echoes Lucretius' notion of \textit{simulacra}, atomic films that peel from objects and mediate perception~\cite{lucretiuscarus1977NatureThings}.  We operationalize this idea by treating biosignals as material casts of cognition.  Rather than representing organoid activity, \sm re-materializes it through agents, light, and clay, asserting an ethics of care in which data co-evolves with its environment.

\sm extends authors' earlier installation \textit{Organoid Protonoesis}~\cite{manoudaki2025OrganoidProtonoesis}, which sonified spontaneous brain-organoid firings as audiovisual transformations. While the prior piece foregrounded representation, the present work slows and streams the same \ch{archival} dataset\footnote{\ch{We use pre-recorded organoid activity for ethical and logistical reasons and will explore live coupling in future work.}} in real time, transforming neurons from data sources into \emph{environmental forces} that steer artificial swarms. Termite builders, slime-mold foragers, and flocking boids thus evolve in direct dialogue with living neural rhythms.

\textbf{Relational Cognition: }  Our design philosophy builds on scholarship that frames cognition as \emph{networked and co-emergent}.  Sheldrake's ``relational consciousness'' positions fungal mycelium as a decentralized intelligence~\cite{sheldrake2021EntangledLifeHow}, while Kac argues that bio-art is fundamentally mutational and co-creative~\cite{kac2007SignsLifeBio}.  Latour's actor-network theory (ANT) likewise distributes agency across assemblages of humans and non-humans, where artefacts can ``authorize, allow, afford, encourage, [or] forbid'' action~\cite[p.72]{latour2005ReassemblingSocialIntroduction}. \sm literalizes these ideas by letting neurons, algorithms, clay, light, and plants participate as equal actants in a sensory field.

\textbf{Emergent Form: }  Complementing this cognitive frame, we draw on morphogenetic theories of form.  D'Arcy Thompson's \emph{On Growth and Form} proposes that biological structure arises from physical forces rather than top-down plans~\cite{thompson1992GrowthForma}.  Weinstock extends this view to architecture, describing buildings as adaptive ecologies~\cite{weinstock2010ArchitectureEmergenceEvolution}, while Batty models cities as self-organising agent systems~\cite{batty2007CitiesComplexityUnderstanding}.  Alexander emphasises contextual ``fit'' achieved through iterative transformations~\cite{alexander2002NotesSynthesisForm}.  By coupling organoid spikes to termite, slime-mold, and flocking agents, \sm operationalizes these principles: global patterns emerge from local interactions modulated by living neural dynamics.

Our work intervenes at this space in three levels. In terms of visualization, existing bio-art seldom integrates high-density neural data with large-scale agent-based simulations in real time. In the realm of sonification, few installations acoustically embody biosignals through physical resonators tuned to neuronal topology, and regarding cyber-physical orchestration, current frameworks rarely synchronize AI image synthesis, GPU-accelerated a-life, and hardware actuation over network protocols. 

The paper makes the following contributions:
\begin{enumerate}
  \item A real-time pipeline that maps 131-channel spike data onto more than 50 million GPU agents with three behavioral models (stigmergic termites, physarum foragers, flocking boids), and AI-assisted diffusion imagery;
  \item A cyber-physical sound system wherein 27 solenoids strike morphogenic ceramic vessels, creating a material sonification of backbone neurons;
  \item A deployment architecture that synchronizes TouchDesigner, Unity, Processing, Max/MSP, and IoT devices via OSC and MQTT with frame-accurate timing;
  \item \ch{A qualitative discussion on how notions of distributed creative agency, emergent expressive complexity, and collective care emerge within \sm's relational components.}
\end{enumerate}

The remainder of the paper is organized as follows: \cref{sec:related-work} surveys related work on collective intelligence and emergence; \cref{sec:system-design} details the data pipeline, physical fabrication, and cyber-physical components; 
\ch{\cref{sec:discussion} discusses the system with lenses from creative agency, complexity, and collective care;}
 and \cref{sec:conclusion} concludes with future directions toward live organoid integration.

\section{Related Work}
\label{sec:related-work}
The installation emerges within a broader field of contemporary practice exploring the entanglement of artificial life, biological systems, responsive environments, and more-than-human aesthetics. Rather than aligning with a single disciplinary tradition, it weaves together insights from generative media art, architectural theory, bio-art, and systems thinking, creating a hybrid ecology of influences.



\subsection{Collective Intelligence in Bio-Art}
Early discourses on collective cognition can be traced back to Aristotle, who proposed that techn\=e ``completes what nature cannot'' by extending purposeful action beyond the organic realm~\cite{aristotleTechnologyNature}. Contemporary bio-art advances this continuity by framing living tissues, microbes, and plants as co-authors of creative systems.

Kac's \textit{Genesis} (1999) encodes biblical text into bacterial DNA that participants mutate remotely, relocating authorship to a micro-ecological network~\cite{kac2020bioart}.  Ben-Ary's \textit{CellF} (2015) cultivates neural stem cells into an ``external brain'' that improvises on a modular synthesizer, foregrounding non-human agency.  Anker's \textit{Astroculture} (2015) and the synthetic-biology project \textit{E.~chromi} (2009) further expand intelligence to plants and engineered bacteria, respectively, revealing how sensing, computation, and adaptation can operate across species boundaries~\cite{anker2021epistemic,todorovic2020reimagining}. 

Despite this rich repertoire, few works draw on the high-dimensional dynamics
of \emph{brain organoids}.  Where existing pieces often treat biological data
as symbolic content, \sm treats neuronal firings as a structuring force that
modulates behaviour across digital, physical, and botanical media.

\subsection{Emergence and Artificial-Life Systems}
Artificial-life (A-life) art investigates how complex patterns arise from local
rules.  Latham and Todd's \textit{Mutator} (1987) evolved biomorphic forms
through interactively guided genetic algorithms~\cite{todorovic2020reimagining}, while Lomas' \textit{Cellular Forms} (2014) generates morphogenesis via particle dynamics~\cite{lomas2014CellularFormsArtistic}. \ch{Wakefield and Ji's \textit{Entanglement} (2025) merges biologically inspired dynamic simulations with non-narrative spatial storytelling to establish artificial ecosystems of forest and fungal networks~\cite{haru2025WeareEntanglement}.} More recently, Wu \textit{et al.}~\cite{wu2024SurveyRecentPractice} surveyed the field of A-life art and found that most works focus on the evolution of agents rather than the more complex dynamics of brain organoids.

These artistic explorations parallel theoretical work on stigmergy and swarm
intelligence. Langton's ``life-as-it-might-be'' suggests that simulating processes like reproduction and metabolism unveils new life forms and sparks creative potentials~\cite{langton1993ArtificialLife}. He further investigates creative potential at the ``edge of
chaos'', where systems hover between order and turbulence~\cite{langton1990ComputationEdgeChaos}.  Penny analyzes emergent art as coordination through environmental traces~\cite{penny2015EmergenceAgencyInteraction}, and Jones models Physarum networks to reveal self-organizing transport efficiency~\cite{jones2010CharacteristicsPatternFormation}. Author's prior work has also explored the edge of chaos with Jones' model in the context of emergent art~\cite{toka2021edge,toka2022edge}.

Within data-art, however, large-scale agent simulations rarely receive continuous biological input. \sm bridges this gap by injecting real-time organoid spikes into three A-life paradigms (termites, slime molds, and boids) thereby fusing living data with GPU-accelerated emergence.

\subsection{Cyber-Physical Ecosystems}
Responsive environments often combine sensing, projection, and actuation yet remain limited to either digital or material domains. Beesley's \textit{Hylozoic Ground} deploys kinetic membranes that respond to touch~\cite{beesley2010HylozoicGroundLiminal}, while Wakefield and Ji's \textit{Infranet} evolves artificial creatures across a city infrastructure dataset~\cite{wakefield2019InfranetGeospatialDataDriven}.  Still, multisensory orchestration that includes sound, light, clay acoustics, and living plants is uncommon.

\sm contributes a unified architecture that synchronizes AI diffusion imagery, multichannel audio, solenoid-struck ceramics, and a forest of live tropical plants via OSC and MQTT.  By embedding organoid rhythms in this hybrid fabric, we extend prior cyber-physical installations toward a genuinely multi-species, multi-modal ecosystem.


\section{System Design}
\label{sec:system-design}
The system is composed of four interwoven layers: the physical environment \ch{(\cref{sec:physical})}, cyber-physical interfaces \ch{(\cref{sec:cyber-physical})}, digital behavioral systems \ch{(\cref{sec:visualComponent})}, sonic landscape and interaction model \ch{(\cref{sec:audio})}. These layers do not operate in isolation; they are mutually responsive and driven by the brain organoid data \ch{(\cref{sec:data-processing})}, forming a coherent ecology where perception, computation, and matter co-create a continuously evolving experience.


\subsection{Human Brain Organoids: Source and Processing}
\label{sec:data-processing}

The installation is driven by a dataset of spontaneous neuronal activity recorded from human brain organoids from the Kosik Neurobiology Lab~\cite{sharf2022FunctionalOrganoidDynamics}. These organoids were cultivated from human induced pluripotent stem cells (iPSCs) and interfaced with high-density CMOS microelectrode arrays capable of capturing single-unit activity across thousands of electrodes. Raw extracellular data were sampled at 20 kHz, enabling precise resolution of action potentials and inter-spike intervals across large neuronal populations. Spike sorting using the Kilosort2 algorithm yielded 131 active neuron channels per organoid~\cite{ vandermolen2023ProtosequencesBrainOrganoids}, which serve as the primary dataset for this system.



This neuronal population has a subset of 27 units, referred to as backbone neurons, that demonstrated consistent temporal coordination and strong interactivity. While the full set of 131 neurons informs the real-time behavior of the digital simulation, only the backbone subset is mapped onto the solenoid component of the cyber-physical system explained in \cref{sec:cyber-physical}: each of the 27 neurons corresponds to an individual solenoid. These physical components pulse and resonate in synchrony with the \ch{pre-recorded} neuron firing data, adding an embodied sonic and kinetic layer to the installation.

In the visual component explained in \cref{sec:visualComponent}, the organoid data are read as a continuous stream, with each of the 131 digital agents in our simulation environment corresponding to one neuron in the dataset. These agents exhibit emergent behavior, such as trail deposition or directional shifts, in direct response to firing events, allowing biological activity to shape the visual and spatial dynamics of the digital system.

\subsection{Physical Components}
\label{sec:physical}

\begin{figure}[htb]
 \centering
 \includegraphics[width=\columnwidth]{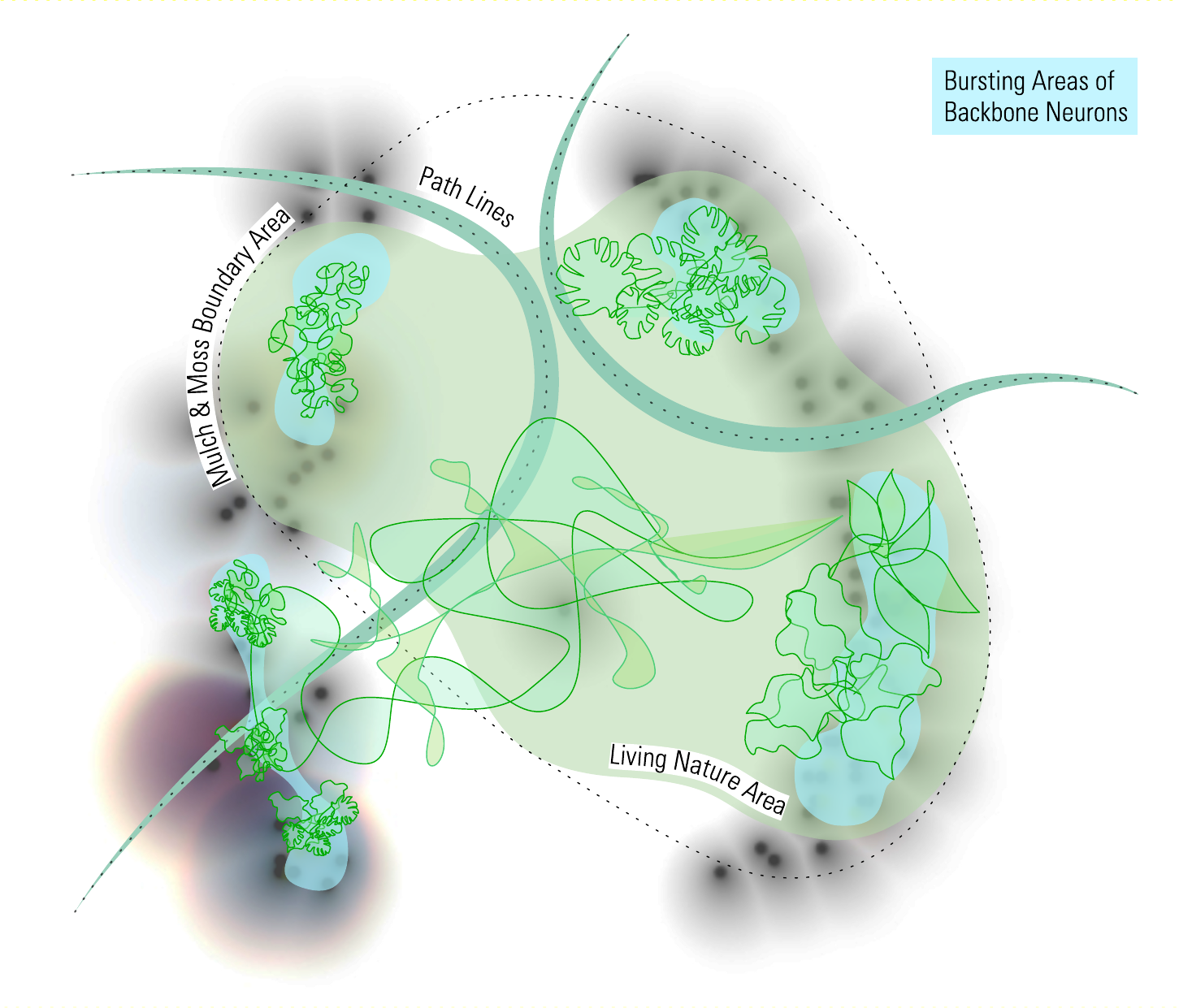}
 \caption{Spatial arrangement diagram. The planning of the forest environment reflected the firing regions of the backbone neurons.}
 \label{fig:spatialArrangement}
\end{figure}

The spatial organization of the forest-like environment was directly informed by the neural structure of the brain organoid itself (\cref{fig:spatialArrangement}). Specifically, we extracted an abstracted topological map from the firing regions of backbone neurons --clusters of coordinated neural activity that formed functional neighborhoods within the dataset. By calculating the average zones where these backbones formed dense interactions, we derived a set of planar paths and region boundaries, which were then used to shape the layout of plant placement and movement corridors within the installation. 

This forest-like environment invites visitors into an immersive terrain shaped by neural rhythm and organic materiality. The planting includes large-leaf species such as Monstera deliciosa, Alocasia, and Strelitzia alba, alongside various hydroponically cultivated plants such as Dracaena sanderiana (bamboos) and Monstrea adansonii (monkey mask). These organisms inhabit a soft substrate composed of mulch and artificial moss, chosen not only for its visual resemblance to undergrowth but also for its olfactory and tactile presence, grounding the digital systems in a material field of scent and softness.

\begin{figure}[htb]
 \centering
 \includegraphics[width=\columnwidth]{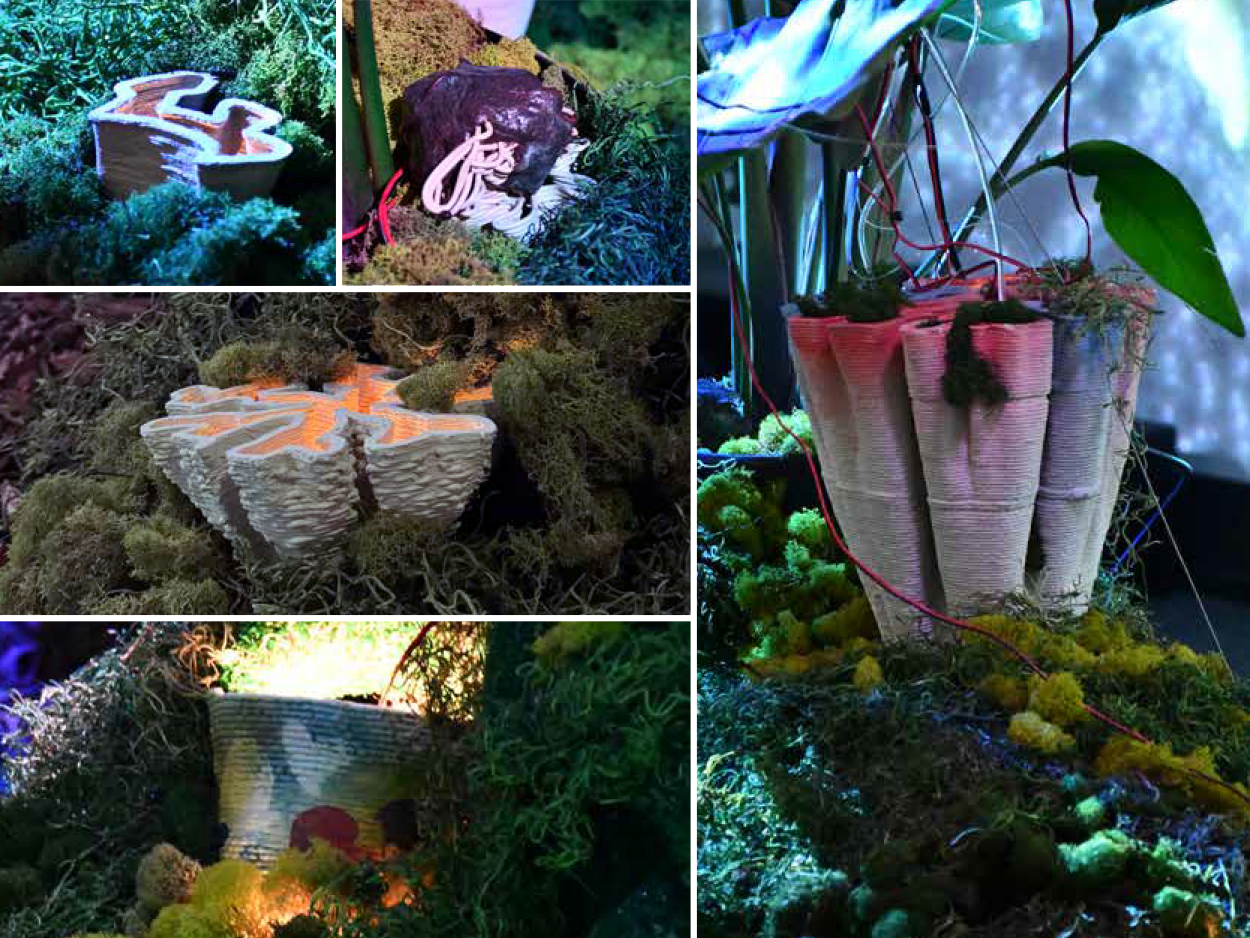}
 \caption{\ch{Photos of five ceramic objects of various sizes (5--60 cm) fabricated using a custom differential growth-based algorithm.}}
 \label{fig:physicalComponents}
\end{figure}

Within this landscape are digitally fabricated morphogenic ceramic vessels of varying dimensions, resembling natural forms like coral reefs (\cref{fig:physicalComponents}). The fabrication process of these objects follows a manual-computational workflow~\cite{toka2023AdaptableWorkflowManualComputational}, in which the goal is to retain manual qualities of making while employing computational techniques for mapping increased complexity with precision onto the material production. While the software guides the vessel's form through extrusion paths, clay's `active' agency~\cite{bennett2010VibrantMatter,ingold2009TextilityMaking} made visible by material memory, drying, and firing characteristics reassert their influence and destabilizes purely digital interventions.
This dialogue enables a deeper engagement between fabrication workflows and care~\cite{toka2024PracticedrivenSoftwareDevelopment}; for the constraints and affordances of clay as a material, for the traditions and tacit knowledge of the craft, for the tools that translate code to form, and for the human-machine partnerships that's mediated in a distributed network of creative agencies.

These tactile artifacts used in \sm have been designed with a custom generative fabrication software developed with Python in Grasshopper/Rhino environment. Unlike the traditional CAD-based fabrication workflows that rely on solid modeling and slicing, these CAM-based toolpaths~\cite{bourgault2023CoilCAMEnablingParametric} have been generated through a rule-based, differential growth algorithm that echo the formative forces of nature~\cite{toka2024CraftingComputationalArtistic}. Upon the formation of the generative toolpath, we engage with traditional ceramic pipeline to prepare the raw stoneware clay bodies, print using a Potterbot clay 3D printer, and post-process the green ware (drying, trimming, bisque firing, glazing, and final firing) to turn them into a finished ceramic artifact.


\subsection{Cyber-Physical Components} 
\label{sec:cyber-physical}
The cyber-physical layer transforms abstract neural rhythms into tactile, acoustic, and luminous phenomena distributed throughout the environment. These components do not simply render data; they respond, resonate, and participate in the ongoing emergence of the system.

\begin{figure}[htb]
 \centering
 \includegraphics[width=\columnwidth]{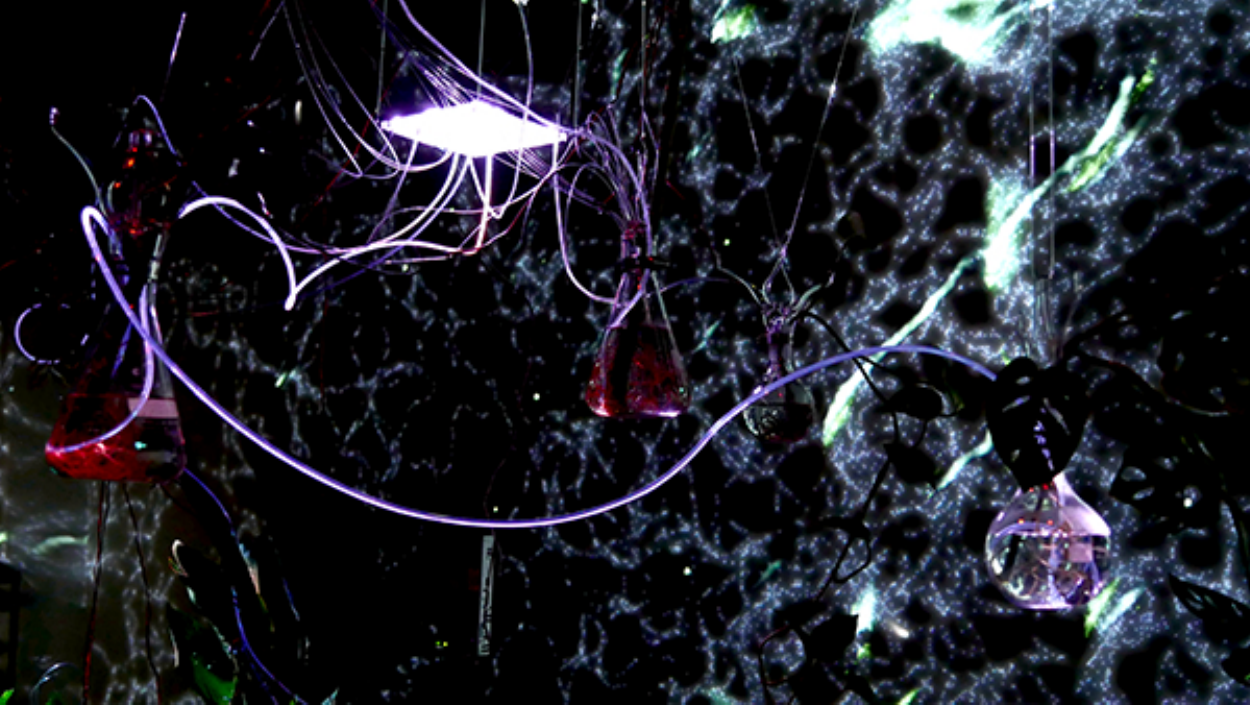}
 \caption{Photo of the LED matrix that drives the cyber-physical components such as fiber optics and solenoids. }
 \label{fig:cyberphysicalComponents}
\end{figure}

Solenoids are embedded in and around the computationally fabricated ceramic vessels and the suspended glassware. Driven by fluctuations in organoid firing intensity, these actuators strike the ceramics with precise timing. The result is a real-time acoustic landscape shaped by collective behavior: a language of micro-strikes, tremors, and pulses that make invisible data physically audible. The vessels are not passive; they resonate differently depending on form, material thickness, and humidity, creating unpredictable feedback. This positions sound not as output but as co-expression of organic and artificial systems.

Two 16x16 RGB LED matrices embedded in custom designed frames drive fiber optics that are blend with the rest of the environment and introduce a stark contrast between the physical and the cyber entities (\cref{fig:cyberphysicalComponents}). The two LED matrices represent the high-density CMOS microelectrode arrays used in the neurobiology lab for organoid data acquisition. These lights pulse and flicker according to real-time agent clustering, neural event intensities, and environmental transitions. Fiber optic strands, embedded in the glassware that carry the hydro plants, carry light into the edges of the installation, creating shimmering micro-landscapes that shimmer in synchrony with neural variation.

Together, these components form what architectural theorist Philip Beesley might call a liminal responsive architecture, a membrane between sensing and sensing back. They allow for material intelligence to emerge, not through AI, but through real-time physical response. The space becomes a soft machine, where matter and signal become entangled in continuous negotiation.


\subsection{Digital Components}
\label{sec:visualComponent}
\begin{figure*}[hbt]
 \centering
\includegraphics[width=\linewidth]{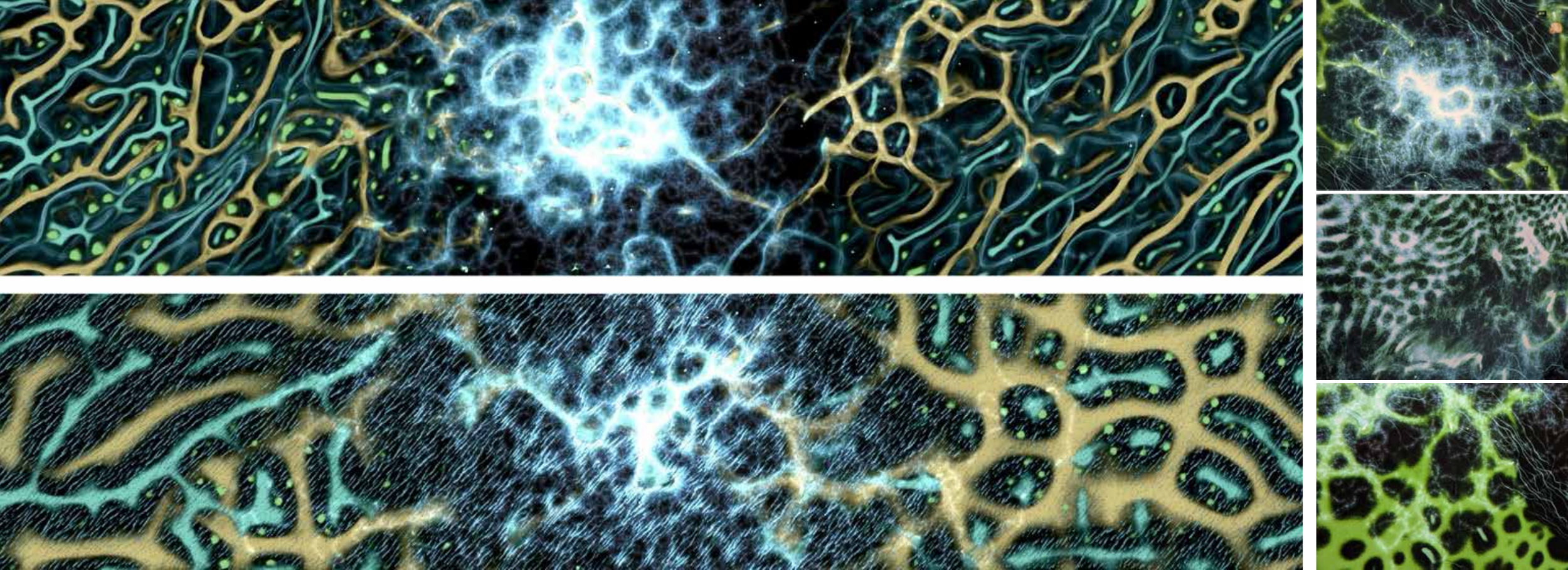}
 \caption{\ch{The artificial visual ecosystems combining termites, slime molds, and flocking agents. (Left) Two high-resolution exports of projected visuals. (Right) Three close-up photos of projection surface showing the variations in a-life simulations with various parameters.}}
 \label{fig:tiffExport}
\end{figure*}

The digital layer (\cref{fig:tiffExport}) forms a dialogue between the physical and cyber-physical components by constructing an ecosystem of various simulated life-like environments. This digital ecosystem is projected on and around the physical components, framing the neural activity. 


The termite simulation is implemented as an agent model and is grounded in principles of self-organization and decentralized coordination. Each agent, corresponding to one of the 131 neurons identified in the spike-sorted organoid dataset, follows a set of local behavioral rules including stochastic motion, environmental sensing, and conditional trail deposition. These agents interact with a scalar field environment that retains memory through dynamic marking and decay, functioning similarly to pheromone trails. Through this feedback-rich loop, the system demonstrates emergent behaviors such as clustering, trail formation, and spatial patterning, all arising without centralized control. Under default conditions, agents operate autonomously within this framework. However, real-time firing events from the organoid dataset directly intervene in the simulation logic: when a neuron exhibits a spike, the corresponding agent deposits a trail at that instant, overriding its endogenous behavior. In this way, trail marking becomes a shared parameter activated either through local environmental feedback or through exogenous biological signals. The system thus integrates spontaneous neural activity into a self-organizing simulation, modulating agent behavior and collective spatial dynamics. The model builds on the conceptual frameworks of Mitchel Resnick's Turtles, Termites, and Traffic Jams, which illustrates emergent order through simple local interactions~\cite[pp.~23-34]{resnick2000TurtlesTermitesTraffic}, and Camazine \textit{et al.}'s Self-Organization in Biological Systems, particularly their articulation of stigmergy, feedback, and randomness as mechanisms underpinning swarm intelligence~\cite[Ch.~5-7)]{camazine2003SelforganizationBiologicalSystems}. 

Building upon this model of decentralized behavior, the slime mold (\textit{Physarum polycephalum}) simulation introduces another self-organizational layer in the digital component. The system implementation uses Jones' modeling of physarum networks~\cite{jones2010CharacteristicsPatternFormation} in a custom compute shader environment in Unity. The slime mold simulations in \sm form a synthetic ecosystem where virtual physarum agents deposit chemical signals (pheromones) and respond to existing signals in the environment. Each agent navigates this dynamic pheromone landscape using a stigmergy communication, and they turn toward denser regions and leave chemical traces that affect the future behaviors. This simple rule-based behavior, combined with the massively parallel structure of GPU programming and increased complexity, allow for producing vast varieties of complex, emergent, and life-like forms that made possible with agentic interactions. Due to the linear time complexity of stigmergy approach ($\mathcal{O}(n)$); where \texttt{n} is the number of agents) and its parallel implementation, the system is capable of simulating a total of 60 million agents split in four species in real-time on a high-end commercial graphics card without throttling the rest of the computations. In the installation, pre-recorded brain organoid signals modulate primary parameters such as sensor angle, sensor distance, movement speed, and turn angle, resulting in continuously shifting, biologically conditioned trail structures. Moreover, these visual traces responds to other a-life simulations discusses in this section. As such, the slime mold simulations enact care as attentiveness: they amplify patterns of distributed responsiveness and mutual shaping, illustrating how cognition might emerge not from isolated brains, but from agent collectives embedded in complex, fluid terrains. 

Another a-life layer is the flocking systems, evoking the collective movement of birds or fish. These entities sense and respond to their neighbors in a direct-communication paradigm through proximity and alignment without a central control. Built upon Reynolds' classical boid model~\cite{reynolds1987FlocksHerdsSchools} and extended through parameter modulation by organoid activity, these simulations demonstrate how emergent coordination can arise from mass applications  of simple interaction rules. Implemented in compute shaders in Unity, each agent reacts to forces such as cohesion, separation, and alignment. The system is capable of simulating 50 thousand agents without jeopardizing the real-time calculations due to quadratic time complexity, $\mathcal{O}(n^2)$, of the direct-communication paradigm. In our system, digital agents adjust their velocity and direction in real-time based on signals derived from \ch{pre-recorded} neuronal populations, effectively becoming extensions of the organoid's temporal rhythms. This results in fluid visual behaviors that echo biological swarm intelligence. 



\subsection{Audio Component}
\label{sec:audio}
The sound environment for \sm is a 16.2-channel generative composition implemented in Max/MSP, driven by neural activity data obtained from brain organoids. The system draws upon principles from the discipline of sonification, employing techniques for translating non-audio data into perceptually meaningful auditory forms that preserve and highlight the dynamic structures inherent in the source signal~\cite{hermann2011SonificationHandbook}. Two primary datasets structure the sonic behavior of the installation: \textit{population firing rate} and \textit{the temporal boundaries of bursting events}. Within the generative system, population firing rate governs the synthesis of parameters and is mapped to determine the dynamic behavior of the soundscape. Moreover, the endpoints of the bursting events were extracted and used to cue harmonic transitions within the composition. 

The harmonic foundation of the work is a progression in C minor with prominent Phrygian inflections. The primary sampled sound material consists of recordings of distorted electric guitar, with each pitch of the progression captured individually across 52 discrete audio files. These files are cued independently by the generative system, enabling dynamic shifts in sonic density and harmonic emphasis. Various treatments are applied to the sampled guitar tones, including granulation~\cite{roads2004Microsound}, playback speed manipulation (resulting in pitch shifting), amplitude envelope shaping, feedback delay, spatial distribution, and reverb. These treatments produce two principal sonic textures: sustained guitar tones and clouds of granulated fragments.

The density of sustained guitar tones is determined by the population firing rate, mapped through a square root function. This results in sparse activity at low firing rates, with event density accelerating nonlinearly toward an upper limit as neural activity increases (range: 0.21–20.0 events/second). A 50\% probability of half-speed playback introduces lower octave variants with longer durations. Each tone has a triangular amplitude envelope with randomly generated attack and decay times, processed through a feedback delay with per-event randomized delay time, and assigned to a random spatial channel.
    
Conversely, granulated guitar textures exhibit behavior inversely related to the firing rate. Here, the firing rate is passed through a sixth root function and inverted, producing dense textures at low rates that thin out as activity intensifies (range: 5.0–160.0 events/second). Grain duration is linearly correlated with firing rate (range: 6.25ms–400ms), with shorter grains reading samples more quickly and thus generating upward pitch shifts. All grain durations are quantized to maintain alignment with harmonic overtones. Each grain is assigned a triangular envelope, with attack and decay shaped by the firing rate processed through a square root function. Spatialization is randomized per grain.

Additional sound sources include sine tone drones and a synthesized low-frequency kick. Sine tones are freely cued at random intervals, each with long attack, sustain, and decay phases, and spatialized independently. The kick is routed exclusively through subwoofers and functions as a low-end anchor. Its event density is inversely mapped to the firing rate (range: 0.1–0.83 events/second), with random chances of producing repeated articulations that suggest fleeting metric structures. The pitch of the kick is chosen randomly between C (the tonal center) and D$\flat$ (Phrygian $\flat$2).

Each of the 16 output channels features its own dedicated reverb processor, contributing to the immersive spatial character of the installation and modulating all audio routed through it.


\subsection{System Synchronization and Data Distribution}

We implemented an application in TouchDesigner to parse and synchronize the neural recording data from the Kosik Neurobiology Lab. The dataset contains 180,000 rows, each representing a millisecond in a 3-minute recording. To enhance perceptibility, we slowed the playback rate to span 90 minutes instead of three. TouchDesigner functions as the master clock, sequentially reading rows and broadcasting a row index via OSC to all connected subsystems.

\begin{figure}[htb]
 \centering
 \includegraphics[width=\columnwidth]{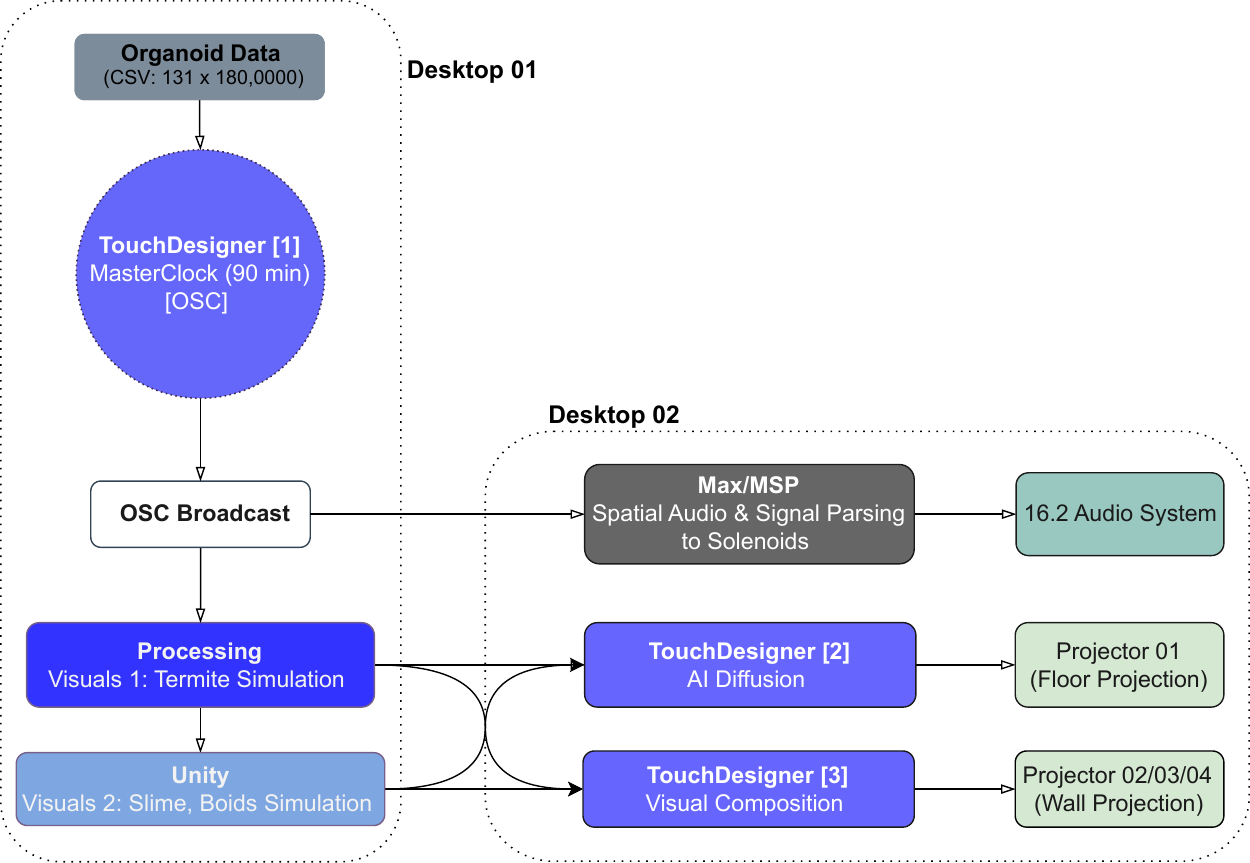}
 \caption{Interaction diagram of the system. Organoid data is parsed in the first TouchDesigner patch that transmits the data through OSC and MQTT. The OSC data is captured by Unity, Processing, second TouchDesigner patch running  Generative AI visuals, and the Max/MSP patch governing the soundscape.}
 \label{fig:interactingComponents}
\end{figure}

This index drives the generative behavior of both visual and physical elements. Unity and Processing handle real-time visualizations on the surrounding walls, 
while a second TouchDesigner process runs a real-time AI diffusion pipeline combining text-to-image and image-to-image generation. This projection is mapped onto the central forest structure. The diffusion model operates using vector images that define the spatial boundaries of the three plant ``islands'' on the ground, alongside text prompts that dynamically control visual output. We used Stable Diffusion 2.1 fine-tuned with a Low-Rank Adaptation (LoRA) model trained on microscopy imagery. The final composite input to the AI model includes the floor plan outlines of the forest and the (x, y) layout of organoid firing positions, which modulate the diffusion input dynamically. 
Max/MSP handles spatialized audio and the solenoid-actuated glassware, both of which respond dynamically to the incoming index value. Both datasets--population firing rate and burst event markers-- were encoded as .wav files and accessed via dataframe indices. These indices were transmitted over a local area network using OSC~\cite{wright1997OpenSoundControlNew} from a TouchDesigner patch that concurrently drives the visual component of the installation. This data-sharing architecture ensures real-time synchronization between audio and visual modalities. 

This architecture enables precise temporal coordination across heterogeneous media outputs, ensuring that all elements respond in synchrony to the unfolding neural dynamics. The pipeline is also network-enabled through Message Queuing Telemetry Transport (MQTT), allowing data to be distributed to remote locations, enabling synchronization across multiple installations.

\subsection{Technical Deployment}
The installation was deployed in the TransLab as part of Deep Cuts, End of Year Show, 2025. The system consisted of two high-performance desktop workstations, each equipped with an NVIDIA RTX 4090 GPU and connected via a 10 Gbps network switch. One machine was responsible for real-time inference of the AI diffusion model, TouchDesigner-based control logic, sound through MAX/MSP, and the cyber-physical system with the solenoids and the fiber optics. The second managed real-time visual rendering using Unity and Processing.

The AI-generated visuals were projected onto the forest structure using a 4K laser floor projector connected to the first machine. The second machine drove a three-projector 4K laser setup arranged in a U-shape configuration, covering two walls, with a combined resolution of 10,184 × 2,160 pixels. Projection alignment and warping were performed using TouchDesigner's native tools. Frame-accurate synchronization across subsystems was achieved using OSC, while Network Device Interface (NDI) was employed to stream textures between the two machines.



\section{\ch{Discussion}}
\label{sec:discussion}
\ch{%
%
%
In this discussion, we reflect on the relational, emergent nature of human--nonhuman collaboration in contemporary media installations. We consider three lenses in \sm: distributed creative agency, emergent expressive complexity, and more-than-human care, and describe how each is put in action in design and operational principles.

\subsection{Distributed Creative Agency}
The human, computational, material, and biological components of \sm all contribute to the overall aesthetics of the installation in various ways. In this distributed setting, creative capacities arise from how partial agencies are connected and how influences are routed between them~\cite{latour2005ReassemblingSocialIntroduction}. Simondon's individuation complements this distributed creative agency perspective: technical objects co-evolve with their milieu, and capacity is realized in coupling rather than in isolated parts~\cite{simondon2011ModeExistenceTechnical}. This also aligns with accounts of material vitality that locate agency in assemblages rather than substances~\cite{bennett2010VibrantMatter}.

In practice, we choose which actants to include, define interfaces and thresholds, and allow organoid activity, simulations, and materials to lead at different moments. This stance aligns with the new materialist notions of guiding material forms without imposition~\cite{ingold2013MaterialsLife} and decentralized coordination in computation~\cite{resnick2000TurtlesTermitesTraffic}. In this configuration, creative agency emerges by the relations themselves, and the artwork is the coordinated pattern of action the network achieves at runtime. We refine it by reconfiguring associations rather than prescribing fixed outputs.

\subsection{Operationalizing Emergent Expressive Complexity}
Within these distributed agency networks, adaptable aesthetic structures emerge from local interactions and feedback. This resonates with how complex systems enable emergent behavior without central control. 
Rather than adapting the quantitative methods of complexity science, we adopt a qualitative artistic stance inspired by the edge of chaos intuition~\cite{langton1990ComputationEdgeChaos}, tuning for transitional regimes where structure can appear under constraints.

Formally, the system can be described as a graph of actants connected by mappings and feedback channels. Edge weights represent coupling strength. A synchronized clock keeps modalities aligned so that changes propagate consistently. Material elements provide memory and transformation. Under these conditions, coherent patterns arise without central control, consistent with the notions of biological self-organization~\cite{camazine2003SelforganizationBiologicalSystems}.
Concretely, a neuronal burst increases termite deposition, Physarum follows enhanced trails, flock cohesion responds, the audio engine shifts texture, solenoids strikes vessels, fiber optics brighten, and visitors perceive the co-occurrences and reorient attention. No single part specifies the whole.

Because agency is distributed, changing thresholds, mappings, or topology produces new yet characteristic behaviors. We evaluate and refine the work by adjusting these couplings for coherence and robustness across modalities, not by scripting final forms. In this sense, the artwork is the emergent behavior of the coupled agencies realized in time.

\subsection{More-Than-Human Care in a Hybrid Ecology}
We approach care as a set of more-than-human and relational practices that support the capacities of living organisms, materials, simulated ecologies, and data subjects~\cite{haraway2016StayingTrouble}. In \sm, care is expressed through design constraints and operational procedures.

For the living elements, we cultivate plants with routine tending through species selection, propagation, misting, cleaning, irrigation, and environmental control. Placement was low-impact and reversible during installation, operation, and teardown. The layout avoids invasive sensing and maintains safe circulation around living elements.

For craft and materials, we work in collaboration with craft communities and align making with the tendencies of clay and the constraints of fabrication. Following Ingold's idea of ``joining forces'' with materials, forms are negotiated rather than imposed~\cite{ingold2013MaterialsLife}. Components are designed for reuse across iterations to reduce waste.

For simulated ecologies, we treat artificial-life layers as systems to steward, not instruments to overdrive. We avoid parameter mappings that override endogenous dynamics. We apply guardrails such as decay, diffusion, and rate limits to prevent collapse or runaway dominance. We prioritize diversity and stability over spectacle and tune timescales for interpretability, aligning with responsive environmental practices that emphasize reciprocity over control~\cite{beesley2010HylozoicGroundLiminal}.

For organoid data, we use pre-recorded activity; no live tissue is present. We map aggregate features such as rates and burst boundaries and align with current characterizations of organoid activity as exhibiting oscillatory dynamics and local synchrony without warranting cognitive attributions~\cite{sharf2022FunctionalOrganoidDynamics} to avoid making anthropomorphic claims. Provenance and consent are tracked by the originating lab. Our use emphasizes transparency through documented mappings, interpretive restraint, and one-way influence only: there is no closed-loop stimulation back to living tissue. In exhibited deployments, we include a concise wall text disclosing the use of archived recordings and the rationale for this choice.

For visualization practice, these measures frame care as more-than-human stewardship. We aim to make relationships legible without extraction and to prefer reversible, maintainable configurations. The goal is not only to represent a multispecies field but to grow it responsibly.

}

\section{Conclusion \& Future Work}
\label{sec:conclusion}
\sm invites a rethinking of what it means to co-create with data, matter, and life. Rather than framing the brain organoid activity as input to be decoded, the installation treats neural signals as part of a responsive ecosystem. Through agent-based simulations, acoustic ceramic vessels, generative soundscape and cyber-physical lighting distributed across a forest-like environment, the system grounds cognition in distributed, embodied, and multisensory experience. Drawing on ecological, cybernetic, and bio-art frameworks, this work challenges anthropocentric models of intelligence and offers an alternate imagery. In this model, form arises from negotiation and mutual influence among human, nonhuman, digital, and organic agents.



In future iterations, we aim to extend this system toward real-time integration with live organoid data streams, moving beyond pre-recorded datasets. By developing a platform that connects directly to in-vitro neural interfaces, we hope to support fully dynamic, bi-directional interactions between living neuronal cultures and computational ecologies. This will allow us to explore how co-authored aesthetic processes evolve under continuous, mutual feedback, opening new pathways for experimental media, artificial life, and embodied data ethics. 
\ch{In parallel, we plan an optional participatory layer that allows audiences to modulate constrained parameters (e.g., mapping gains or diffusion rates) without overriding endogenous neural dynamics, improving legibility while preserving the work's ethics of care.}

\acknowledgments{
 \ch{The authors wish to thank the Kosik Neurobiology Lab at UC Santa Barbara, Professor Kenneth Kosik, and Tjitse van der Molen, Ph.D., as well as Professor Marcos Novak and the transLAB at the Media Arts and Technology program, for enabling a sincere collaborative environment that supports integration of art and science. We also thank UC Santa Barbara's Expressive Computation Lab, NSF IIS Future of Work at the Human-Technology Frontier Program (Award: 2026286) and the NSF IIS Human-Centered Computing Program (Award: 2007094) for supporting the clay fabrication research used in this work.}
}

\bibliographystyle{abbrv-doi}

\bibliography{references}

\begin{thebibliography}{10}

\bibitem{alexander2002NotesSynthesisForm}
C.~Alexander.
\newblock {\em Notes on the Synthesis of Form}.
\newblock Harvard Univ. Press, 17. printing ed., 2002.

\bibitem{anker2021epistemic}
S.~Anker.
\newblock Epistemic practices in bio art.
\newblock 36(6):1389--1394, 2021-11-01. doi: {{%
10\hspace{.1pt}\discretionary{.}{%
}{.}\hspace{.4pt}1007\discretionary{/}{%
}{/}s00146\discretionary{%
}{-}{-}021\discretionary{%
}{-}{-}01152\discretionary{%
}{-}{-}w}}


\bibitem{aristotleTechnologyNature}
Aristotle.
\newblock Aristotle on technology and nature, 1999.
\newblock Translated excerpts and commentary compiled by Joachim Schummer.

\bibitem{batty2007CitiesComplexityUnderstanding}
M.~Batty.
\newblock {\em Cities and Complexity: Understanding Cities with Cellular Automata, Agent-Based Models, and Fractals}.
\newblock MIT, 1. paperback ed ed., 2007.

\bibitem{beesley2010HylozoicGroundLiminal}
P.~Beesley, H.~Isaacs, P.~Ohrstedt, and R.~Gorbet, eds.
\newblock {\em Hylozoic Ground: Liminal Responsive Architecture: {{Philip Beesley}}}.
\newblock Riverside Architectural Press, first edition ed., 2010.

\bibitem{bennett2010VibrantMatter}
J.~Bennett.
\newblock {\em Vibrant {{Matter}}}.
\newblock Duke University Press, 2010. doi: {{%
10\hspace{.1pt}\discretionary{.}{%
}{.}\hspace{.4pt}2307\discretionary{/}{%
}{/}j\hspace{.1pt}\discretionary{.}{%
}{.}\hspace{.4pt}ctv111jh6w}}


\bibitem{bourgault2023CoilCAMEnablingParametric}
S.~Bourgault, P.~Wiley, A.~Farber, and J.~Jacobs.
\newblock {{CoilCAM}}: {{Enabling Parametric Design}} for {{Clay 3D Printing Through}} an {{Action-Oriented Toolpath Programming System}}.
\newblock In {\em Proceedings of the 2023 {{CHI Conference}} on {{Human Factors}} in {{Computing Systems}}}, {{CHI}} '23, pp. 1--16. Association for Computing Machinery, 2023-04-19. doi: {{%
10\hspace{.1pt}\discretionary{.}{%
}{.}\hspace{.4pt}1145\discretionary{/}{%
}{/}3544548\hspace{.1pt}\discretionary{.}{%
}{.}\hspace{.4pt}3580745}}


\bibitem{camazine2003SelforganizationBiologicalSystems}
S.~Camazine, ed.
\newblock {\em Self-Organization in Biological Systems}.
\newblock Princeton Studies in Complexity. Princeton Univ. Press, 2. print., and 1. paperback print ed., 2003.

\bibitem{haraway2016StayingTrouble}
D.~J. Haraway.
\newblock {\em Staying with the Trouble: Making Kin in the Chthulucene}.
\newblock Duke University Press, Durham, NC, 2016.

\bibitem{hermann2011SonificationHandbook}
T.~Hermann, A.~Hunt, and J.~G. Neuhoff, eds.
\newblock {\em The Sonification Handbook}.
\newblock Logos Verlag, 2011.

\bibitem{ingold2009TextilityMaking}
T.~Ingold.
\newblock The textility of making.
\newblock 34(1):91--102, 2009. doi: {{%
10\hspace{.1pt}\discretionary{.}{%
}{.}\hspace{.4pt}1093\discretionary{/}{%
}{/}cje\discretionary{/}{%
}{/}bep042}}


\bibitem{ingold2013MaterialsLife}
T.~Ingold.
\newblock The {{Materials}} of {{Life}}.
\newblock In {\em Making: {{Anthropology}}, {{Archaeology}}, {{Art}} and {{Architecture}}}, pp. 17--31. Routledge, 2013.

\bibitem{haru2025WeareEntanglement}
H.~H. Ji and G.~Wakefield.
\newblock Entanglement: an immersive art of an engagement with non-conscious intelligence.
\newblock In S.~W. Roh and Y.~H. Roh, eds., {\em ISEA2025: 30th International Symposium on Electronic Art – Exhibition Catalogue}. Art Center Nabi, 2025.

\bibitem{jones2010CharacteristicsPatternFormation}
J.~Jones.
\newblock Characteristics of {{Pattern Formation}} and {{Evolution}} in {{Approximations}} of {{Physarum Transport Networks}}.
\newblock 16(2):127--153, 2010-04. doi: {{%
10\hspace{.1pt}\discretionary{.}{%
}{.}\hspace{.4pt}1162\discretionary{/}{%
}{/}artl\hspace{.1pt}\discretionary{.}{%
}{.}\hspace{.4pt}2010\hspace{.1pt}\discretionary{.}{%
}{.}\hspace{.4pt}16\hspace{.1pt}\discretionary{.}{%
}{.}\hspace{.4pt}2\hspace{.1pt}\discretionary{.}{%
}{.}\hspace{.4pt}16202}}


\bibitem{kac2007SignsLifeBio}
E.~Kac.
\newblock {\em Signs of Life: Bio Art and Beyond}.
\newblock Leonardo. the MIT press, 2007.

\bibitem{kac2020bioart}
E.~Kac.
\newblock Bio art.
\newblock 36(6):1367--1376, 2020-11-01. doi: {{%
10\hspace{.1pt}\discretionary{.}{%
}{.}\hspace{.4pt}1007\discretionary{/}{%
}{/}s00146\discretionary{%
}{-}{-}020\discretionary{%
}{-}{-}00958\discretionary{%
}{-}{-}4}}


\bibitem{langton1990ComputationEdgeChaos}
C.~G. Langton.
\newblock Computation at the edge of chaos: {{Phase}} transitions and emergent computation.
\newblock 42(1--3):12--37, 1990-06. doi: {{%
10\hspace{.1pt}\discretionary{.}{%
}{.}\hspace{.4pt}1016\discretionary{/}{%
}{/}0167\discretionary{%
}{-}{-}2789\discretionary{%
}{(}{(}90\discretionary{)}{%
}{)}90064\discretionary{%
}{-}{-}V}}


\bibitem{langton1993ArtificialLife}
C.~G. Langton.
\newblock Artificial {{Life}}.
\newblock In {\em {{ARS Electronica Catalog}}}, 1993.

\bibitem{latour2005ReassemblingSocialIntroduction}
B.~Latour.
\newblock {\em Reassembling the Social: An Introduction to Actor-Network-Theory}.
\newblock Clarendon Lectures in Management Studies. Oxford University Press, Oxford ; New York, 2005.

\bibitem{lomas2014CellularFormsArtistic}
A.~Lomas.
\newblock Cellular forms: An artistic exploration of morphogenesis.
\newblock In {\em {{ACM SIGGRAPH}} 2014 {{Studio}} on - {{SIGGRAPH}} '14}, pp. 1--1. ACM Press, Vancouver, Canada, 2014. doi: {{%
10\hspace{.1pt}\discretionary{.}{%
}{.}\hspace{.4pt}1145\discretionary{/}{%
}{/}2619195\hspace{.1pt}\discretionary{.}{%
}{.}\hspace{.4pt}2656282}}


\bibitem{lucretiuscarus1977NatureThings}
T.~Lucretius~Carus.
\newblock {\em The Nature of Things}.
\newblock Norton, 1st ed ed., 1977.

\bibitem{manoudaki2025OrganoidProtonoesis}
N.~Manoudaki, I.~Paterakis, D.~Flatley, R.~Millett, and M.~Novak.
\newblock Organoid\_protonoesis ii.
\newblock In S.~W. Roh and Y.~H. Roh, eds., {\em ISEA2025: 30th International Symposium on Electronic Art – Exhibition Catalogue}, pp. 54--55. Art Center Nabi, 2025.

\bibitem{penny2015EmergenceAgencyInteraction}
S.~Penny.
\newblock Emergence, {{Agency}}, and {{Interaction}}—{{Notes}} from the {{Field}}.
\newblock 21(3):271--284, 2015-08-01. doi: {{%
10\hspace{.1pt}\discretionary{.}{%
}{.}\hspace{.4pt}1162\discretionary{/}{%
}{/}ARTL\_a\_00167}}


\bibitem{resnick2000TurtlesTermitesTraffic}
M.~Resnick.
\newblock {\em Turtles, Termites, and Traffic Jams: Explorations in Massively Parallel Microworlds}.
\newblock Complex Adaptive Systems. MIT Press, 6. printing ed., 2000.

\bibitem{reynolds1987FlocksHerdsSchools}
C.~W. Reynolds.
\newblock Flocks, {{Herds}}, and {{Schools}}: {{A Distributed Behavioral Model}}.
\newblock 21(4):25--34, 1987-08. doi: {{%
10\hspace{.1pt}\discretionary{.}{%
}{.}\hspace{.4pt}1145\discretionary{/}{%
}{/}37402\hspace{.1pt}\discretionary{.}{%
}{.}\hspace{.4pt}37406}}


\bibitem{roads2004Microsound}
C.~Roads.
\newblock {\em Microsound}.
\newblock MIT Press, 1. paperback ed ed., 2004.

\bibitem{sharf2022FunctionalOrganoidDynamics}
T.~Sharf, T.~van~der Molen, S.~M.~K. Glasauer, E.~Guzman, A.~P. Buccino, G.~Luna, Z.~Cheng, M.~Audouard, K.~G. Ranasinghe, K.~Kudo, S.~S. Nagarajan, K.~R. Tovar, L.~R. Petzold, A.~Hierlemann, P.~K. Hansma, and K.~S. Kosik.
\newblock Functional neuronal circuitry and oscillatory dynamics in human brain organoids, 2022-07-29. doi: {{%
10\hspace{.1pt}\discretionary{.}{%
}{.}\hspace{.4pt}1038\discretionary{/}{%
}{/}s41467\discretionary{%
}{-}{-}022\discretionary{%
}{-}{-}32115\discretionary{%
}{-}{-}4}}


\bibitem{sheldrake2021EntangledLifeHow}
M.~Sheldrake.
\newblock {\em Entangled Life: {{How Fungi Make}} Our {{Worlds}}, {{Change}} Our {{Minds}} \& {{Shape}} Our {{Futures}}}.
\newblock Random House, random house trade paperback edition ed., 2021.

\bibitem{simondon2011ModeExistenceTechnical}
G.~Simondon.
\newblock {\em On the Mode of Existence of Technical Objects}.
\newblock Univocal Publishing, Minneapolis, MN, 2017.

\bibitem{thompson1992GrowthForma}
D.~W. Thompson.
\newblock {\em On Growth and Form}.
\newblock Dover Publications, the complete revised edition ed., 1992.

\bibitem{todorovic2020reimagining}
V.~Todorovic.
\newblock Reimagining life (forms) with generative and bio art.
\newblock 36(6):1323--1329, 2020-03-04. doi: {{%
10\hspace{.1pt}\discretionary{.}{%
}{.}\hspace{.4pt}1007\discretionary{/}{%
}{/}s00146\discretionary{%
}{-}{-}020\discretionary{%
}{-}{-}00937\discretionary{%
}{-}{-}9}}


\bibitem{toka2021edge}
M.~Toka.
\newblock The edge of chaos.
\newblock In {\em Hybrid Science Experimentation}. EXP. Experimental Photo Festival, Barcelona, Spain (online), 2021.
\newblock Group exhibition curated by Felicita Russo and Maciej Zapiór.

\bibitem{toka2022edge}
M.~Toka.
\newblock The edge of chaos.
\newblock In {\em SYMADES '22}. California NanoSystems Institute, Santa Barbara, CA, USA, 6 2022.
\newblock Group exhibition curated by Marko Peljhan.

\bibitem{toka2024CraftingComputationalArtistic}
M.~Toka.
\newblock Crafting the {{Computational}}: {{Artistic Production}}, {{Generative Systems}}, and {{Digital Fabrication}}.
\newblock In {\em Designing {{Interactive Systems Conference}}}, pp. 24--29. ACM, 2024-07. doi: {{%
10\hspace{.1pt}\discretionary{.}{%
}{.}\hspace{.4pt}1145\discretionary{/}{%
}{/}3656156\hspace{.1pt}\discretionary{.}{%
}{.}\hspace{.4pt}3665122}}


\bibitem{toka2023AdaptableWorkflowManualComputational}
M.~Toka, S.~Bourgault, C.~Friedman-Gerlicz, and J.~Jacobs.
\newblock An {{Adaptable Workflow}} for {{Manual-Computational Ceramic Surface Ornamentation}}.
\newblock In {\em Proceedings of the 36th {{Annual ACM Symposium}} on {{User Interface Software}} and {{Technology}}}, {{UIST}} '23, pp. 1--15. Association for Computing Machinery, 2023-10-29. doi: {{%
10\hspace{.1pt}\discretionary{.}{%
}{.}\hspace{.4pt}1145\discretionary{/}{%
}{/}3586183\hspace{.1pt}\discretionary{.}{%
}{.}\hspace{.4pt}3606726}}


\bibitem{toka2024PracticedrivenSoftwareDevelopment}
M.~Toka, D.~Frost, S.~Bourgault, A.~Farber, C.~Friedman-Gerlicz, R.~Lee, E.~Paek, P.~Wiley, and J.~Jacobs.
\newblock Practice-driven {{Software Development}}: {{A Collaborative Method}} for {{Digital Fabrication Systems Research}} in a {{Residency Program}}.
\newblock In {\em Designing {{Interactive Systems Conference}}}, pp. 1192--1217. ACM, 2024-07. doi: {{%
10\hspace{.1pt}\discretionary{.}{%
}{.}\hspace{.4pt}1145\discretionary{/}{%
}{/}3643834\hspace{.1pt}\discretionary{.}{%
}{.}\hspace{.4pt}3661522}}


\bibitem{vandermolen2023ProtosequencesBrainOrganoids}
T.~Van Der~Molen, A.~Spaeth, M.~Chini, S.~Hernandez, G.~A. Kaurala, H.~E. Schweiger, C.~Duncan, S.~{McKenna}, J.~Geng, M.~Lim, J.~Bartram, A.~Dendukuri, Z.~Zhang, J.~Gonzalez-Ferrer, K.~Bhaskaran-Nair, L.~J. Blauvelt, C.~R. Harder, L.~R. Petzold, D.-M. Alam El~Din, J.~Laird, M.~Schenke, L.~Smirnova, B.~M. Colquitt, M.~A. Mostajo-Radji, P.~K. Hansma, M.~Teodorescu, A.~Hierlemann, K.~B. Hengen, I.~L. Hanganu-Opatz, K.~S. Kosik, and T.~Sharf.
\newblock Protosequences in brain organoids model intrinsic brain states, 2023-12-30. doi: {{%
10\hspace{.1pt}\discretionary{.}{%
}{.}\hspace{.4pt}1101\discretionary{/}{%
}{/}2023\hspace{.1pt}\discretionary{.}{%
}{.}\hspace{.4pt}12\hspace{.1pt}\discretionary{.}{%
}{.}\hspace{.4pt}29\hspace{.1pt}\discretionary{.}{%
}{.}\hspace{.4pt}573646}}


\bibitem{wakefield2019InfranetGeospatialDataDriven}
G.~Wakefield and H.~H. Ji.
\newblock Infranet: A geospatial data-driven neuro-evolutionary artwork.
\newblock In {\em 2019 {IEEE} {VIS} Arts Program ({VISAP})}, pp. 1--7. {IEEE}, 2019-10. doi: {{%
10\hspace{.1pt}\discretionary{.}{%
}{.}\hspace{.4pt}1109\discretionary{/}{%
}{/}VISAP\hspace{.1pt}\discretionary{.}{%
}{.}\hspace{.4pt}2019\hspace{.1pt}\discretionary{.}{%
}{.}\hspace{.4pt}8900903}}


\bibitem{weinstock2010ArchitectureEmergenceEvolution}
M.~Weinstock.
\newblock {\em The Architecture of Emergence: The Evolution of Form in Nature and Civilisation}.
\newblock Wiley, 1. publ ed., 2010.

\bibitem{wright1997OpenSoundControlNew}
M.~Wright and A.~Freed.
\newblock Open {{SoundControl}}: {{A New Protocol}} for {{Communicating}} with {{Sound Synthesizers}}.
\newblock In {\em Proceedings of the 1997 {{International Computer Music Conference}}}, pp. 101--104, 1997.

\bibitem{wu2024SurveyRecentPractice}
Z.-W. Wu, H.~Qu, and K.~Zhang.
\newblock A survey of recent practice of artificial life in visual art.
\newblock {\em Artificial Life}, 30(1):106--135, 2024. doi: {{%
10\hspace{.1pt}\discretionary{.}{%
}{.}\hspace{.4pt}1162\discretionary{/}{%
}{/}artl\_a\_00433}}


\end{thebibliography}
\end{document}